\documentclass[]{aastex7}

\usepackage{newtxtext,newtxmath}
\usepackage[T1]{fontenc}
\usepackage[encapsulated]{CJK}

\begin{document}
\begin{CJK*}{UTF8}{gbsn}
\title{Long Photometric Cycles in Double Periodic Variables from Nodal Precession of a Tilted Accretion Disk}

\author[0000-0002-7663-7900,gname='承亮',sname='焦']{Cheng-Liang Jiao (焦承亮)}
\email{jiaocl@ynao.ac.cn}
\affiliation{Yunnan Observatories, Chinese Academy of Sciences, 396 Yangfangwang, Guandu District, Kunming, 650216, People's Republic of China}
\affiliation{Center for Astronomical Mega-Science, Chinese Academy of Sciences, 20A Datun Road, Chaoyang District, Beijing, 100012, People's Republic of China}
\affiliation{Key Laboratory for the Structure and Evolution of Celestial Objects, Chinese Academy of Sciences, 396 Yangfangwang, Guandu District, Kunming, 650216, People's Republic of China} 

\author{Er-gang Zhao}
\email{zergang@ynao.ac.cn}
\affiliation{Yunnan Observatories, Chinese Academy of Sciences, 396 Yangfangwang, Guandu District, Kunming, 650216, People's Republic of China}
\affiliation{Key Laboratory for the Structure and Evolution of Celestial Objects, Chinese Academy of Sciences, 396 Yangfangwang, Guandu District, Kunming, 650216, People's Republic of China}
\author[0000-0002-0796-7009]{Liying Zhu}
\email{zhuly@ynao.ac.cn}
\affiliation{Yunnan Observatories, Chinese Academy of Sciences, 396 Yangfangwang, Guandu District, Kunming, 650216, People's Republic of China}
\affiliation{University of Chinese Academy of Sciences, No. 1 Yanqihu East Road, Huairou District, 101408, Beijing, People's Republic of China}
\affiliation{Key Laboratory for the Structure and Evolution of Celestial Objects, Chinese Academy of Sciences, 396 Yangfangwang, Guandu District, Kunming, 650216, People's Republic of China}
\author[]{Azizbek Matekov}
\email{azizbek_matekov@mail.ru}
\affiliation{Yunnan Observatories, Chinese Academy of Sciences, 396 Yangfangwang, Guandu District, Kunming, 650216, People's Republic of China}
\affiliation{University of Chinese Academy of Sciences, No. 1 Yanqihu East Road, Huairou District, 101408, Beijing, People's Republic of China}
\affiliation{Ulugh Beg Astronomical Institute, Uzbekistan Academy of Sciences, 33 Astronomicheskaya str., Tashkent, 100052, Uzbekistan}

\correspondingauthor{Cheng-Liang Jiao (焦承亮), Er-gang Zhao}
\email{jiaocl@ynao.ac.cn, zergang@ynao.ac.cn}

\begin{abstract}
We investigate whether the long photometric cycles observed in double-periodic variables (DPVs) can arise from nodal precession of a tilted accretion disk driven by the tidal torque of the companion. Within a simple analytical framework, we derive testable relations linking the long-to-orbital period ratio to the binary mass ratio, the normalized disk size, and the disk tilt angle $\beta$, which itself can be inferred from the long-cycle amplitude, orbital inclination $i$, and disk luminosity fraction. 
The model naturally reproduces the two observed long-cycle light-curve morphologies---sinusoidal and double-hump---distinguished by the geometric criterion $i+\beta \le 90^\circ$ versus $i+\beta>90^\circ$.
Applying these relations to a sample of DPVs, we find that the inferred disk sizes are physically reasonable and consistent with independent light-curve modeling for a non-negligible subset of systems. 
Our results show that tidal nodal precession represents a viable and potentially important contributor to the long-period variability of DPVs and provide a quantitative framework for future observational and theoretical studies.
\end{abstract}

\section{Introduction}\label{intro}

Double-periodic variables (DPVs) are a class of semi-detached interacting binary systems characterized by two distinct photometric cycles: a short period associated with the orbital motion and a long photometric cycle, $P_{\rm long}$, typically an order of magnitude longer than the orbital period, $P_{\rm orb}$ \citep[e.g.,][]{Men2003,Pol2010,Paw2013}.
The class was first recognized in surveys of the Magellanic Clouds and later identified in the Galaxy, with long cycles on average $\sim 33\,P_{\rm orb}$ \citep{Men2010p,Men2017}.
Observationally, the long cycle manifests as a sinusoidal or sometimes double-hump modulation with typical amplitudes of a few tenths of a magnitude \citep{Gar2025}.

Despite extensive observational and modeling efforts, the physical origin of the long photometric cycle in DPVs remains unsettled. Proposed explanations broadly invoke cyclic variations in mass transfer and accretion-related processes, including donor-star magnetic activity, structural changes in a geometrically and optically thick accretion disk, disk--stream interaction regions, winds or episodic mass loss, and variable circumstellar or circumbinary material \citep[e.g.,][]{Lin2006, Men2008, Des2010, Dju2010, Men2012b, Sch2017, Gar2018, Cal2025}.

In this work, we explore an alternative and complementary possibility: that the long photometric cycle in at least a subset of DPVs arises from nodal precession of a tilted accretion disk driven by the tidal torque of the companion. Tilted-disk precession is a well-established phenomenon in other classes of interacting binaries and naturally predicts a long timescale regulated by the orbital period and the system mass ratio \citep[e.g.,][]{Katz1973, Katz1980, Katz1982, Pap1995, Lar1998, Rom2000}. 
If the disk radiates approximately axisymmetrically, the changing projected area of a precessing tilted disk produces a photometric modulation that is inherently sinusoidal or, in certain geometric configurations, double-hump (see Section~\ref{Fmod}), qualitatively reproducing the observed long-cycle light-curve morphologies of DPVs.

In the context of DPVs, nodal precession of a tilted accretion disk was briefly proposed as a possible explanation for the long photometric cycle in AU~Mon by \citet{Viv1998}. Subsequently, \citet{Men2003} expressed skepticism about its applicability, arguing that the apparent clustering of the observed period ratios $P_{\rm long}/P_{\rm orb} \sim 35$ at that time was difficult to reconcile with those inferred for precessing-disk systems identified in X-ray binaries.
However, these early discussions were largely qualitative, and a quantitative framework connecting the nodal precession period to system parameters---such as mass ratio, disk size, and viewing geometry---has been lacking.
With the DPV sample now expanded from $\sim30$ systems \citep{Men2003} to over 200 systems \citep[e.g.,][]{Men2017,Glo2024}, the development of such a framework is timely.

Nevertheless, disk precession is not expected to operate universally across all DPVs. For example, \citet{Wil1999} argued against a nodal precession origin for the long-period variability in the extreme system $\beta$~Lyrae, while \citet{Men2012b} reached a similar conclusion for the DPV V393~Sco.
These results suggest that nodal precession is unlikely to account for the long photometric cycles in certain individual systems. However, given the small number of such cases reported to date, the mechanism could still operate in other DPVs.

In this study, we do not aim to provide a comprehensive census of all DPVs, nor to claim that tilted-disk precession operates in every system. Rather, we aim to derive testable relations for the tilted-disk precession model, including expressions for the long-to-orbital period ratio and the amplitude of the long cycle, and to assess the physical plausibility of this mechanism based on observational constraints.
We note that the present work does not address the physical origin of the disk tilt itself. 
Possible mechanisms---such as stream--disk misalignment, anisotropic mass transfer, or magnetic warping---have been discussed extensively in other classes of interacting binaries. 
Our goal here is to examine whether, once a tilt is present, tidal nodal precession can account for the observed long-cycle phenomenology of DPVs.

The paper is organized as follows. Section~\ref{model} presents the analytical framework, Section~\ref{app} applies it to observed DPVs, and Section~\ref{sum} summarizes our conclusions.

\section{Tilted Disk Precession and Photometric Modulation}\label{model}
\subsection{Tidal nodal precession of a tilted accretion disk}

We consider a semi-detached interacting binary system with orbital separation $a$ and orbital period $P_{\rm orb}$.
The accretor of mass $M_1$ is surrounded by an accretion disk fed by mass transfer from a donor star of mass $M_2$.
The binary mass ratio is defined as $q \equiv M_2/M_1$.
The accretion disk is assumed to be tilted by an angle $\beta$ with respect to the orbital plane.

Under the secular tidal torque exerted by the companion, a tilted disk undergoes retrograde nodal precession.
In the general case, the nodal precession angular frequency can be obtained by integrating the tidal torque over the disk structure \citep{Pap1995},
\begin{equation}\label{omega_p_int}
\Omega_{\mathrm{p}} = -\frac{3}{4} \frac{G M_2}{a^3}
\frac{\int_{R_{\mathrm{in}}}^{R_{\mathrm{d}}} \Sigma r^3 \,\mathrm{d} r}
{\int_{R_{\mathrm{in}}}^{R_{\mathrm{d}}} \Sigma \Omega_{\mathrm{K}} r^3 \,\mathrm{d} r}
\cos \beta ,
\end{equation}
where $R_{\mathrm{in}}$ and $R_{\mathrm{d}}$ are the inner and outer radii of the disk, $\Sigma(r)$ is the surface density profile, and $\Omega_{\mathrm{K}}(r) = \sqrt{GM_1/r^3}$ is the Keplerian angular velocity.
This expression requires detailed knowledge of the disk structure that is not available for most DPVs.

\citet{Rom2000} showed that, for physically reasonable disk parameters the disk-integrated precession frequency approximately reduces to the expression for a rigidly precessing ring \citep{Katz1982}.
In particular, they found that for an adiabatic index $\gamma \simeq 5/3$ the two formulations are in close agreement, while mathematically they become identical for $\gamma = 1.5$.
This result implies that the precessing-ring approximation captures the leading-order behavior of nodal precession for realistic accretion disks.
Therefore, in the absence of detailed constraints on the disk structure, we adopt the precessing-ring expression,
\begin{equation}\label{omega_p}
\Omega_{\mathrm{p}} = -\frac{3}{4} \frac{G M_2}{a^3} \frac{1}{\Omega_{\mathrm{K}}(R_{\mathrm{d}})} \cos \beta
\end{equation}
to evaluate the nodal precession period of the tilted disk in DPVs.
Here $R_{\mathrm{d}}$ represents a characteristic radius at which the tidal torque effectively acts on the disk.
In the context of the disk-integrated formulation, this characteristic radius corresponds to the outer disk radius \citep{Rom2000}. This identification is appropriate as long as the disk precesses approximately as a rigid body, which requires that the sound-crossing timescale across the disk is shorter than the precession timescale \citep{Pap1995}.

\subsection{Photometric modulation from a precessing disk}\label{Fmod}

The observed flux of a DPV system can be decomposed into a stellar component and a disk component,
\begin{equation}\label{F_tot}
F_{\rm tot}(t) = F_{\ast}(t) + F_{\rm disk}(t),
\end{equation}
where $F_{\ast}(t)$ represents the combined contribution from the two stars, including ellipsoidal modulation and eclipses, and $F_{\rm disk}(t)$ denotes the emission associated with the accretion disk.

The stellar flux $F_{\ast}(t)$ varies predominantly on the orbital timescale and can be modeled independently using standard light-curve synthesis techniques.
These short-period variations do not affect the origin or the timescale of the long-period modulation discussed below, which is entirely governed by the disk geometry.

\begin{figure}[ht!]
	\includegraphics[width=\columnwidth]{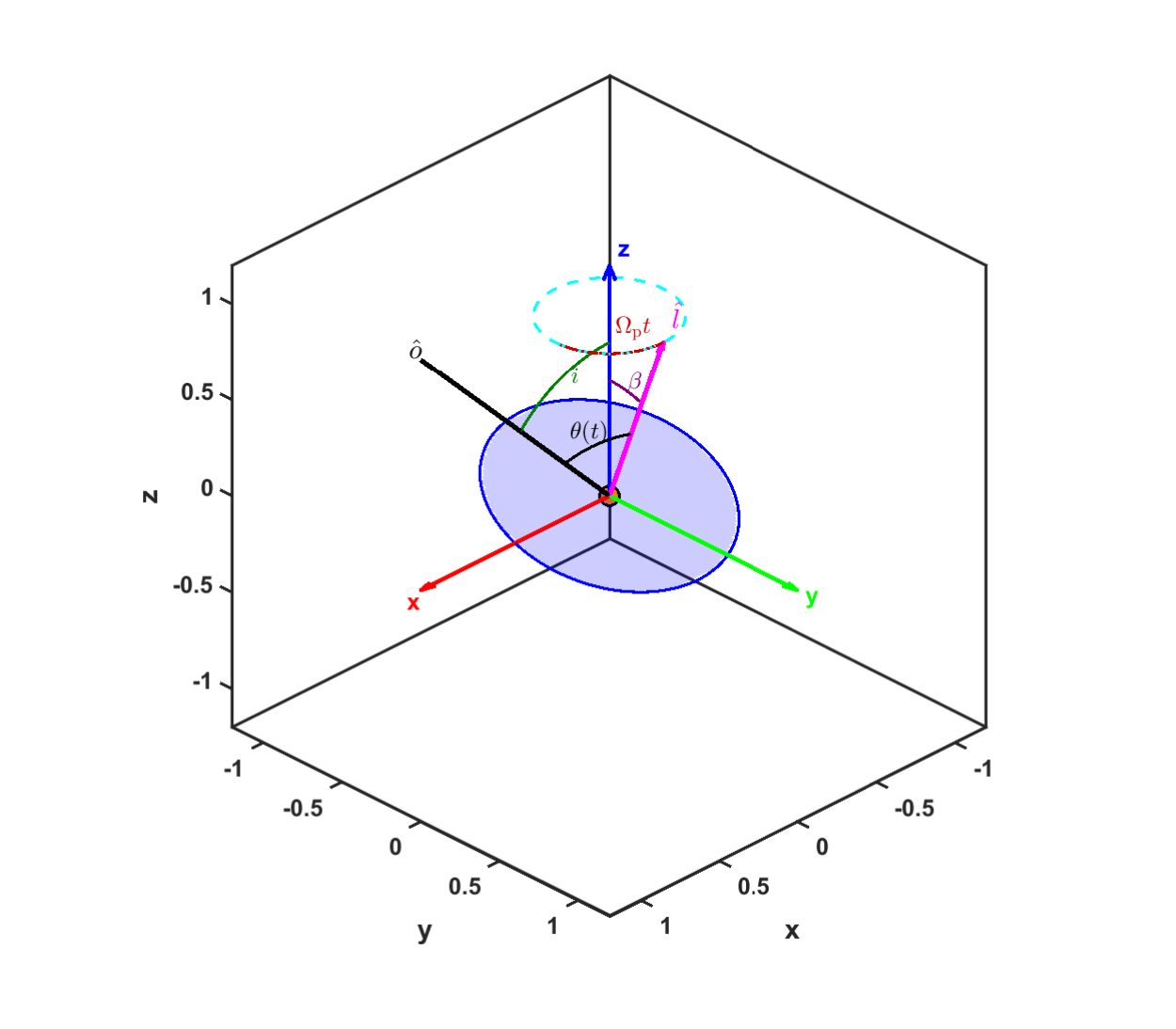}
\caption{Schematic illustration (not to scale) of a tilted accretion disk undergoing nodal precession around the accretor. The accretor is located at the origin, and the binary orbital plane is defined as the $x$--$y$ plane. The observer's line of sight lies in the $x$--$z$ plane and is denoted by the unit vector $\hat{\boldsymbol{o}}$, while the unit vector along the disk angular momentum is denoted by $\hat{\boldsymbol{l}}$. This geometry defines the angles used in the derivation of the photometric modulation.}
	\label{fig1}
\end{figure}

To describe the geometric modulation produced by a precessing disk, we introduce a right-handed Cartesian coordinate system, as illustrated in Figure~\ref{fig1}.
The orbital angular momentum vector is taken to define the $z$-axis, such that the binary orbital plane coincides with the $x$--$y$ plane.
The observer's line of sight lies in the $x$--$z$ plane and makes an inclination angle $i$ with respect to the orbital angular momentum, such that the unit vector pointing toward the observer is
\begin{equation}
\hat{\boldsymbol{o}} = (\sin i,\,0,\,\cos i).
\end{equation}

The unit vector along the disk angular momentum is denoted by $\hat{\boldsymbol{l}}$.
The disk is tilted by an angle $\beta$ relative to the orbital plane and undergoes nodal precession about the $z$-axis with angular frequency $\Omega_{\rm p}$.
Choosing the precession phase such that the disk angular momentum lies in the $x$--$z$ plane at $t=0$, the disk orientation can be written as
\begin{equation}
\hat{\boldsymbol{l}} =
\bigl(
\sin\beta \cos \Omega_{\rm p} t,\,
\sin\beta \sin \Omega_{\rm p} t,\,
\cos\beta
\bigr).
\end{equation}

The instantaneous angle $\theta(t)$ between the disk normal and the observer's line of sight is then given by
\begin{equation}
\cos \theta(t)
= \hat{\boldsymbol{o}} \cdot \hat{\boldsymbol{l}}
= \cos i \cos \beta
+ \sin i \sin \beta \cos(\Omega_{\rm p} t).
\end{equation}

In the simplest geometric limit, the observed disk flux is proportional to the projected  emitting area of the disk, and hence to $\cos \theta(t)$.
The disk contribution to the total flux can therefore be written as
\begin{equation}
F_{\rm disk}(t) = F_0 \cos \theta(t)
= F_0 \bigl[
\cos i \cos \beta
+ \sin i \sin \beta \cos(\Omega_{\rm p} t)
\bigr],
\end{equation}
where $F_0$ is a normalization constant that depends on the intrinsic disk luminosity.

Defining the mean disk flux as
\begin{equation}
\langle F_{\rm disk} \rangle = F_0 \cos i \cos \beta ,
\end{equation}
the disk flux can be expressed as
\begin{equation}\label{F_disk}
F_{\rm disk}(t)
=
\langle F_{\rm disk} \rangle
\left[
1 + \tan i \tan \beta \cos(\Omega_{\rm p} t)
\right].
\end{equation}
This expression describes a purely sinusoidal modulation with angular frequency $\Omega_{\rm p}$, provided that the angle between the disk angular momentum vector and the line of sight remains no more than $90^\circ$ throughout the precession cycle.
This condition is satisfied when $i+\beta \le 90^\circ$.

When $i+\beta>90^\circ$, the instantaneous angle $\theta(t)$ exceeds $90^\circ$ during part of the precession cycle.
In this case, the observed disk flux, which scales with the projected emitting area, is proportional to $|\cos\theta(t)|$ rather than $\cos\theta(t)$.
As a result, the long-cycle modulation exhibits a characteristic double-hump structure within a single precession period.
Importantly, although the waveform contains two maxima per cycle, the fundamental modulation period remains identical to the nodal precession period.

Taken together, this simple geometric consideration naturally and self-consistently reproduces the two basic classes of long-cycle light-curve morphology in DPVs, namely sinusoidal and double-hump modulations, and provides a clear geometric criterion for distinguishing between them.
Because the disk tilt angle $\beta$ is expected to be small in most interacting binaries \citep[e.g.,][]{Ogi2001,Fou2010}, the condition $i+\beta>90^\circ$ is satisfied only in systems viewed at relatively high inclinations.
Consequently, double-hump long-cycle light curves are expected to occur less frequently than sinusoidal ones, in agreement with observational reviews of DPVs \citep[e.g.,][]{Gar2025}.

In realistic systems, the accretion disk possesses a finite vertical thickness.
Geometric effects associated with this thickness, such as self-occultation at the outer rim or obscuration of the inner emitting regions, can modify the detailed waveform of the long-cycle variability.
Similarly, departures from axisymmetry in the disk emissivity can introduce higher-order harmonics superposed on the basic precessional modulation.
While these effects may modify the waveform of the light curve, they do not change the fundamental modulation period.

By contrast, mutual eclipses between the stellar components and the disk occur on the orbital timescale.
Such effects therefore contribute only to short-period variability and can be absorbed into the modeling of the stellar flux term $F_{\ast}(t)$, without affecting the origin or timescale of the long photometric cycle.

Consequently, the fundamental timescale of the long-period variability is set solely by the nodal precession frequency of the tilted disk.
The observed long photometric period in DPVs can therefore be directly identified with the disk precession period,
\begin{equation}
P_{\rm long} = P_{\rm prec} = \frac{2\pi}{|\Omega_{\rm p}|}.
\end{equation}
\subsection{Period ratio and disk size}

The orbital angular frequency of the binary is
\begin{equation}
\Omega_{\rm orb} = \left( \frac{G(M_1+M_2)}{a^3} \right)^{1/2} .
\end{equation}
The ratio between the long period and the orbital period is therefore given by
\begin{equation}\label{period_ratio}
\frac{P_{\rm long}}{P_{\rm orb}}
= \frac{\Omega_{\rm orb}}{|\Omega_{\rm p}|}
=
\frac{4}{3} \frac{\sqrt{1+q}}{q} 
\left( \frac{a}{R_{\mathrm{d}}} \right)^{3/2}
\frac{1}{\cos \beta} .
\end{equation}
Equation~\eqref{period_ratio} directly links the observed long-to-short period ratio in DPVs to the system mass ratio, disk size, and tilt angle.

Solving Equation~\eqref{period_ratio} for the characteristic disk size yields
\begin{equation}\label{eq:rd_over_a}
\frac{R_d}{a}
=
\left[
\frac{4}{3}
\frac{\sqrt{1+q}}{q}
\left( \frac{P_{\rm orb}}{P_{\rm long}} \right)
\frac{1}{\cos\beta}
\right]^{2/3}.
\end{equation}
For typical DPVs, the observed period ratio is $P_{\rm long}/P_{\rm orb} \simeq 33$, 
and mass ratios are usually in the range $q \simeq 0.15$--$0.35$ \citep{Men2017}. 
For a small disk tilt ($\cos\beta \simeq 1$), Equation~\eqref{eq:rd_over_a} implies $R_{\rm d}/a \sim 0.25$--$0.45$, physically reasonable for semi-detached binaries.
\subsection{Disk tilt angle and long-cycle photometric amplitudes}

According to Equation~\eqref{F_disk}, the disk flux varies sinusoidally over a nodal
precession cycle and reaches its extrema at $\cos(\Omega_{\rm p} t)=\pm1$.
The maximum and minimum disk fluxes are therefore given by
\begin{align}
F_{\rm disk}^{\rm max} &=
\langle F_{\rm disk} \rangle
\left( 1 + \tan i \tan \beta \right), \\
F_{\rm disk}^{\rm min} &=
\langle F_{\rm disk} \rangle
\left( 1 - \tan i \tan \beta \right).
\end{align}

The total observed flux (Equation~\eqref{F_tot}) consists of a stellar component
and a disk component.
Since the stellar flux varies predominantly on the orbital timescale, it can be
replaced by its long-cycle average, $\langle F_{\ast} \rangle$, when considering
the long-period modulation.
The maximum and minimum total fluxes over the long cycle are then
\begin{align}
F_{\rm tot}^{\rm max} &=
\langle F_{\ast} \rangle +
\langle F_{\rm disk} \rangle
\left( 1 + \tan i \tan \beta \right), \\
F_{\rm tot}^{\rm min} &=
\langle F_{\ast} \rangle +
\langle F_{\rm disk} \rangle
\left( 1 - \tan i \tan \beta \right).
\end{align}

Following common practice in the DPV literature, the long-cycle photometric
amplitude is defined as the peak-to-peak magnitude variation \citep[e.g.,][]{Pol2010},
\begin{equation}
\Delta V \equiv
V_{\rm min} - V_{\rm max}
=
2.5 \log_{10}
\left(
\frac{F_{\rm tot}^{\rm max}}{F_{\rm tot}^{\rm min}}
\right),
\end{equation}
where $V_{\rm min}$ and $V_{\rm max}$ correspond to the faintest and brightest
states (largest and smallest magnitudes) of the system over the long cycle, respectively.

Defining the disk-to-total mean flux ratio as
\begin{equation}
f_{\rm d} \equiv
\frac{\langle F_{\rm disk} \rangle}{\langle F_{\rm tot} \rangle},
\end{equation}
the photometric amplitude can be written as
\begin{equation}
10^{0.4 \Delta V}=
\frac{1 + f_{\rm d} \tan i \tan \beta}
{1 - f_{\rm d} \tan i \tan \beta}.
\end{equation}

Solving for the disk tilt angle yields
\begin{equation}\label{beta}
\tan \beta=
\frac{10^{0.4 \Delta V} - 1}
{f_{\rm d} \tan i \left( 10^{0.4 \Delta V} + 1 \right)}.
\end{equation}

Equation~\eqref{beta} provides a direct method to infer the disk tilt angle $\beta$ from observables.
The derived $\beta$ can be used as a consistency check in individual systems, for example to assess whether the inferred tilt is reasonable and whether $i+\beta$ predicts the observed long-cycle light-curve morphology.

For systems with $i+\beta>90^\circ$, the disk becomes self-occulted during part of the
precession cycle, such that the minimum projected disk flux is truncated at
$F_{\rm disk}^{\rm min}=0$.
In this case, Equation~\eqref{beta} becomes
\begin{equation}
\tan\beta
=
\frac{(1-f_{\rm d})\,10^{0.4\Delta V}-1}
{f_{\rm d}\tan i},
\end{equation}
which applies to DPVs exhibiting double-hump long-cycle light curves.

%
\section{Application to Observed DPVs}\label{app}
\subsection{Inferred Disk Sizes in DPVs}

We collected 13 DPVs from the literature for which a single study provides measurements of both the accretor and donor masses, as well as the orbital and long photometric periods\footnote{The extreme system $\beta$~Lyrae and the DPV V393~Sco are excluded because previous studies argued against a nodal-precession origin for their long-period variability \citep{Wil1999, Men2012b}. Systems with strongly variable or disappearing long cycles (e.g., OGLE~BLG-ECL-157529 and Au~Mon; \citealt{Men2021,Cel2025}) are also excluded, as a strongly time-dependent precession period is beyond the scope of the simple analytical framework adopted here. By contrast, OGLE-LMC-DPV-65 was retained: although its long cycle decreased in the past, \citet{Men2019} showed that this trend subsequently ceased and the long period had remained nearly constant for about a decade, making it suitable for our analysis.} (Table~\ref{tab1}). 
This selection ensures internal consistency among the parameters adopted in our analysis.
For each system, we compute the dimensionless disk radius $(R_{\rm d}/a)_{\rm mod}$ predicted by the tilted-disk nodal precession model using Equation~\eqref{eq:rd_over_a}.
While the disk tilt angle $\beta$ is unknown for most sources, it is typically expected to be small in interacting binaries with tilted disks \citep{Ogi2001,Fou2010}, and we therefore adopt $\cos\beta \simeq 1$ as a representative value.
An exception is HD~170582, for which published decomposed luminosities of the stars and the disk, together with the long-cycle amplitude and orbital inclination, allow the disk tilt angle to be inferred (Section~\ref{detail}).
We also neglect orbital eccentricity and identify the orbital separation with the semi-major axis.
This is a reasonable approximation for DPVs, which are typically in nearly circular orbits, with reported eccentricities, when available, being at the level of a few percent or less.
Uncertainties are propagated using the values reported in the original references; for quantities without published uncertainties, the corresponding parameters are treated as fixed inputs in the error propagation.

\begin{figure}[ht!]
	\includegraphics[width=\columnwidth]{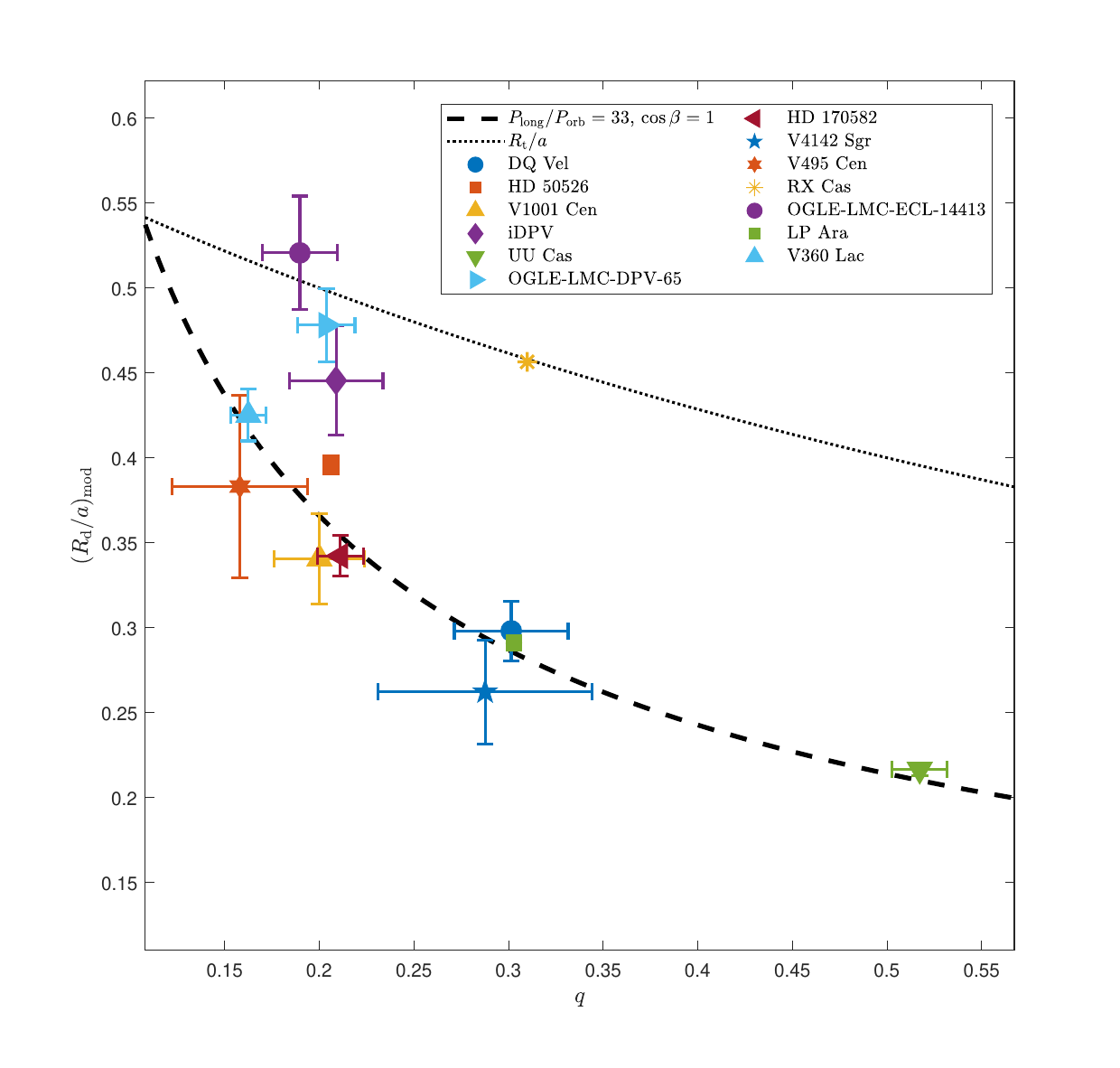}
	\caption{
Model-predicted disk radius normalized by the orbital separation, $(R_{\rm d}/a)_{\rm mod}$, as a function of the mass ratio $q$, for DPVs listed in Table~\ref{tab1}.
The dashed line shows the theoretical relation for a representative period ratio $P_{\rm long}/P_{\rm orb}=33$ with $\cos\beta=1$.
The dotted line indicates the tidal radius $R_{\rm t}/a$.
For some systems, error bars are absent or very small because uncertainties in the relevant input parameters are not reported or are negligible in the original references.
}
	\label{fig2}
\end{figure}

Figure~\ref{fig2} shows $(R_{\rm d}/a)_{\rm mod}$ as a function of the mass ratio $q$ for the DPVs listed in Table~\ref{tab1}.
The dashed line illustrates the theoretical relation for a representative period ratio $P_{\rm long}/P_{\rm orb}=33$ and $\cos\beta=1$.
Deviations from this line reflect differences in the observed period ratios; variations in tilt could introduce additional scatter but are not explored here except for HD~170582. 
The dotted line represents the dimensionless tidal radius,
\begin{equation}
\frac{R_{\rm t}}{a} = \frac{0.6}{1+q},
\end{equation}
which is commonly interpreted as an upper limit on the radial extent of a stable accretion disk in interacting binaries \citep{Pac1977,War1995}.
As shown in Figure~\ref{fig2}, most inferred disk sizes lie below the tidal radius.
The only exception is OGLE-LMC-ECL-14413, which nominally exceeds this limit but remains consistent with it within the quoted uncertainty.
%
\begin{deluxetable*}{lccccccc}
\tablecaption{DPVs with Measured Masses and Photometric Periods
\label{tab1}}
\tablehead{
\colhead{Object} &
\colhead{$M_1$ $(M_{\odot})$} &
\colhead{$M_2$ $(M_{\odot})$} &
\colhead{$P_{\rm orb}$ (d)} &
\colhead{$P_{\rm long}$ (d)} &
\colhead{$(R_{\rm d}/a)_{\rm mod}$} &
\colhead{$(R_{\rm d}/a)_{\rm obs}$} &
\colhead{Ref.\tablenotemark{b}}
}
\startdata
DQ Vel            & $7.3 \pm 0.3$   & $2.2 \pm 0.2$   & $6.083299(7)$    & $188.7(2)$ & $0.298 \pm 0.018$ & $0.434 \pm 0.011$ & 1 \\
HD 50526          & $5.48 \pm 0.02$       & $1.13 \pm 0.02$       & $6.701(1)$       & $191(2)$ & $0.396 \pm 0.005$      & $0.526 \pm 0.001$       & 2 \\
V1001 Cen         & $4.25 \pm 0.10$  & $0.85 \pm 0.10$  & $6.73(1)$        & $247.3(100)$ & $0.341 \pm 0.027$ & $0.512 \pm 0.006$        & 3 \\
iDPV\tablenotemark{a}  & $9.1 \pm 0.5$   & $1.9 \pm 0.2$   & $7.284297(10)$   & $172$    & $0.446 \pm 0.032$ & $0.401 \pm 0.015$ & 4 \\
UU Cas            & $17.4 \pm 0.3$  & $9.0 \pm 0.2$   & $8.519296(8)$    & $268.7(16)$ & $0.216 \pm 0.004$      & $0.399 \pm 0.027$ & 5 \\
OGLE-LMC-DPV-65   & $13.8 \pm 0.3$  & $2.81 \pm 0.20$  & $10.0316267(56)$  & $218$    & $0.478 \pm 0.022$ & $0.501 \pm 0.008$        & 6 \\
HD 170582         & $9.0 \pm 0.2$   & $1.9 \pm 0.1$   & $16.87177(2084)$    & $587$    & $0.342 \pm 0.012$ & $0.340 \pm 0.005$        & 7 \\
V4142 Sgr         & $3.86 \pm 0.30$  & $1.11 \pm 0.20$  & $30.633(2)$      & $1201(14)$ & $0.262 \pm 0.031$ & $0.325 \pm 0.005$        & 8 \\
V495 Cen          & $5.76 \pm 0.30$ & $0.91 \pm 0.20$ & $33.492(2)$      & $1283$   & $0.383 \pm 0.054$ & $0.486 \pm 0.016$ & 9 \\
RX Cas            & $5.81$          & $1.8$           & $32.31190(31)$   & $516$    & $0.457$           & $0.426 \pm 0.014$ & 10 \\
OGLE-LMC-ECL-14413 & $5.8 \pm 0.3$  & $1.1 \pm 0.1$   & $38.15917(54)$   & $778.8(46)$ & $0.521 \pm 0.034$ & $0.421 \pm 0.043$ & 11 \\
LP Ara            & $9.84$          & $2.98$          & $8.53295$        & $273$    & $0.291$           & --              & 12 \\
V360 Lac          & $7.45 \pm 0.30$ & $1.21 \pm 0.05$ & $10.085449(27)$  & $322$    & $0.425 \pm 0.015$ & --              & 13 \\
\enddata
\tablenotetext{a}{Also referred to as 
OGLE J051553.32$-$692558.1.}
\tablenotetext{b}{References: (1) \citet{Bar2013}; (2) \citet{Ros2021}; (3) \citet{Cal2025}; (4) \citet{Gar2013}; (5) \citet{Men2020}; (6) \citet{Men2019}; (7) \citet{Men2015}; (8) \citet{Ros2023}; (9) \citet{Ros2018}; (10) \citet{Men2022}; (11) \citet{Men2025}; (12) \citet{Men2011}; (13) \citet{Lin2006}.}
\tablecomments{
The table is ordered by increasing orbital period, with systems lacking reported disk radii listed at the end.
Uncertainties are given either in parentheses or using the $\pm$ notation when available from the literature; when uncertainties are not reported, none are shown.
The quantity $(R_{\rm d}/a)_{\rm mod}$ is computed from the tilted-disk nodal precession model using the observed periods and masses.
The quantity $(R_{\rm d}/a)_{\rm obs}$ is derived from the disk radius $R_{\rm d}$ and the orbital separation $a$ reported in the references, typically obtained from detailed light-curve modeling, and is listed when available.
}
\end{deluxetable*}
%
%

\begin{figure}[ht!]
	\includegraphics[width=\columnwidth]{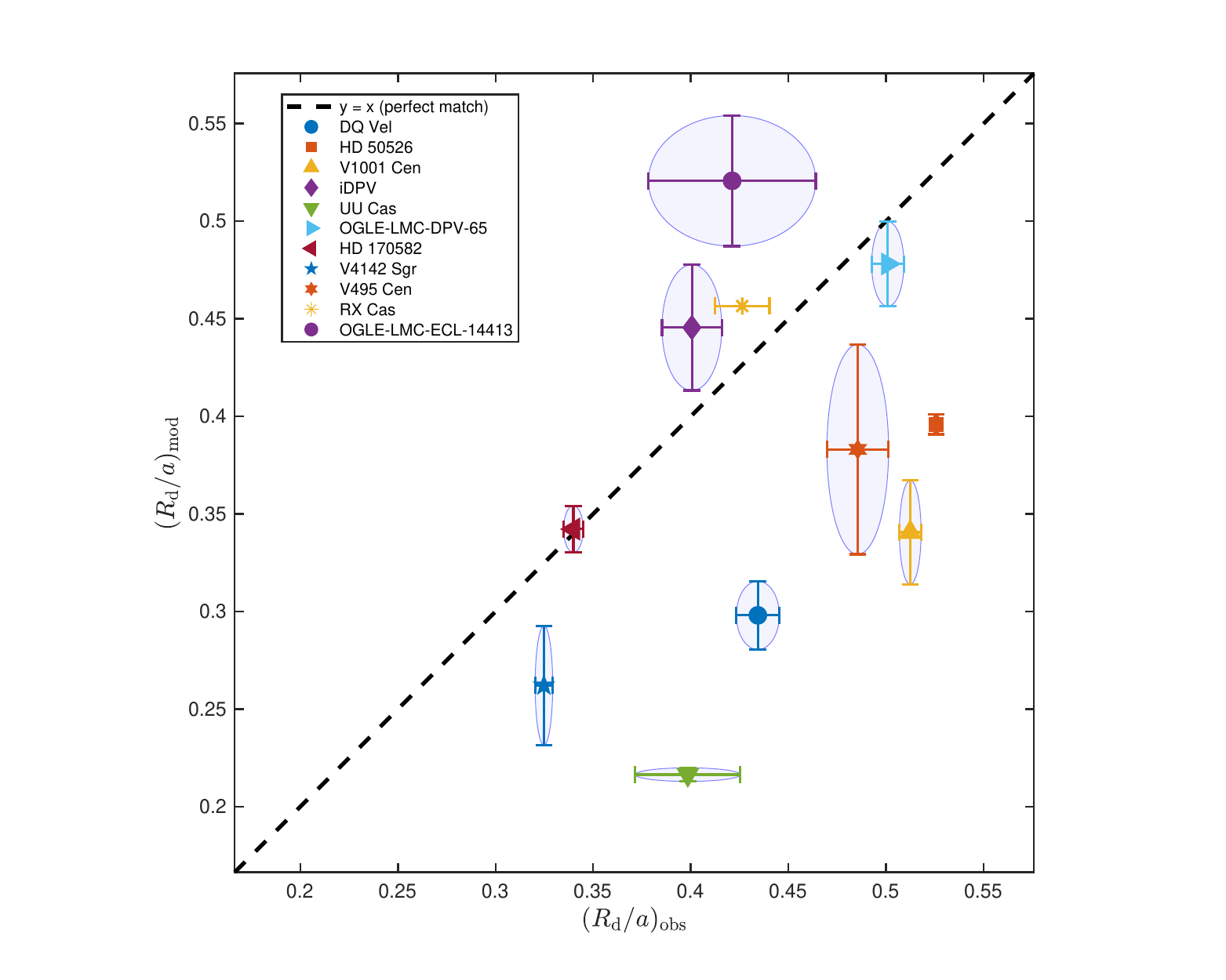}
\caption{
Comparison between the disk radius inferred from the tilted-disk precession model,
$(R_{\rm d}/a)_{\rm mod}$, and the observationally inferred values $(R_{\rm d}/a)_{\rm obs}$.
Each point represents one of the 11 DPVs with available observational disk-radius estimates.
The dashed line indicates equality.
Uncertainty ellipses denote the $1\sigma$ uncertainties of individual systems.
}
	\label{fig3}
\end{figure}

Among the systems listed in Table~\ref{tab1}, eleven have published estimates of both the disk radius $R_{\rm d}$ and the orbital separation $a$, allowing a direct comparison between $(R_{\rm d}/a)_{\rm mod}$ and the observationally inferred ratio $(R_{\rm d}/a)_{\rm obs}$ (Figure~\ref{fig3}). 
Agreement can be assessed by whether the corresponding uncertainty region intersects the $y=x$ line.
%
%

We find that approximately 18\% (2/11) of systems are consistent within $1\sigma$, 45\% (5/11) within $2\sigma$, and 64\% (7/11) within $3\sigma$. 
To further quantify the association between $(R_{\rm d}/a)_{\rm mod}$ and $(R_{\rm d}/a)_{\rm obs}$, we computed both Spearman and Pearson correlation coefficients.
The Spearman rank correlation coefficient $\rho = 0.33$ ($p = 0.33$, $n = 11$) indicates a positive, though not statistically significant, monotonic trend, while the Pearson correlation coefficient $r = 0.36$ ($p = 0.28$) shows a similar pattern for linear association. 
Given the small sample size, the lack of statistical significance is not unexpected. 
Bootstrap resampling ($n = 10{,}000$) yields a 95\% confidence interval of $[-0.38, 0.77]$ for $\rho$, indicating substantial uncertainty but remaining consistent with a positive association.

Although a monotonic trend is present, the scatter in Figure~\ref{fig3} is substantial, indicating that the model does not provide a one-to-one prediction of the observationally inferred disk sizes. 
This dispersion likely reflects a combination of factors, including the possibility that long cycles in some systems are driven by mechanisms other than tilted-disk nodal precession, the use of literature input parameters without any fine-tuning to match individual disk sizes, the heterogeneous origin of the observational disk-radius estimates relying on different modeling assumptions and error treatments, intrinsic deviations from the rigid-disk approximation, and temporal variability in disk structure.

Overall, the tilted-disk precession model reproduces the inferred disk sizes for a non-negligible subset of the systems in our sample, supporting its role as a viable and potentially important contributor to the long-cycle phenomenology of DPVs. 
At the same time, the substantial dispersion in Figure~\ref{fig3} highlights the need for caution when applying the model to individual systems.

\subsection{A Detailed Consistency Check: HD~170582}\label{detail}

HD~170582 is the only system in our sample with published, quantitatively decomposed
luminosities of both the stellar components and the accretion disk, together with
measurements of the long-cycle photometric amplitude $\Delta V$ and the orbital
inclination $i$.
These quantities enable an independent estimate of the disk tilt angle $\beta$ using Equation~\eqref{beta}, providing a stringent consistency check of the tilted-disk precession framework.

Specifically, the primary-star luminosity is $2858 \pm 529\,L_{\odot}$, the secondary-star luminosity is $863 \pm 80\,L_{\odot}$, and the accretion-disk luminosity is $1564\,L_{\odot}$.\footnote{The original reference, \citet{Men2016}, quotes a typical accuracy of $<10\%$ for the disk luminosity, but no numerical uncertainty is provided. 
For our calculation we conservatively treat it as a fixed value, effectively corresponding to the lower bound of the observational uncertainty.}
The orbital inclination is $i = 67.4^\circ \pm 0.4^\circ$, and the amplitude of the long-period photometric variability is $\Delta V = 0.1$~mag.
All these quantities are derived from observational data or observation-constrained modeling \citep{Men2015,Men2016}.
Substituting them into Equation~\eqref{beta}, we obtain a disk tilt angle of
$\beta = 3.7^\circ \pm 0.4^\circ$,
where the uncertainty is dominated by the uncertainty in the disk-to-total luminosity ratio.

Although most quantitative constraints on disk tilt angles come from studies of X-ray binaries and cataclysmic variables, these systems share the same essential physical ingredients---mass transfer, tidal torques, and viscous accretion disks---as DPVs.
In this broader context of interacting binaries, the inferred tilt angle is well within the range expected from previous studies \citep[e.g.,][]{Ogi2001,Fou2010}.
The obtained $i+\beta = 71.1^\circ \pm 0.8^\circ < 90^\circ$, indicating a sinusoidal rather than double-hump modulation, consistent with the observed long-cycle light curve of HD~170582.

Substituting $\beta = 3.7^\circ \pm 0.4^\circ$ and the system parameters of HD~170582 listed in Table~\ref{tab1} into Equation~\eqref{eq:rd_over_a}, we obtain $(R_{\rm d}/a)_{\rm mod} = 0.342 \pm 0.012$.
By contrast, detailed light-curve modeling yields a disk radius of $R_{\rm d} = 20.8 \pm 0.3\,R_\odot$ and an orbital separation of $a = 61.2 \pm 0.2\,R_\odot$ \citep{Men2015}, corresponding to
$(R_{\rm d}/a)_{\rm obs}=0.340\pm0.005$.
The two estimates agree within $1\sigma$, which is noteworthy given the simplicity of the tilted-disk nodal precession model and the fact that all input parameters are taken directly from published observational constraints, without any additional fitting or fine-tuning.

Finally, we assess whether the accretion disk in HD~170582 can plausibly undergo approximately rigid nodal precession, as assumed in our analytical framework.
Adopting the observationally constrained disk-edge temperature
$T_{\rm d} \simeq (5.4$--$5.7)\times10^{3}$~K \citep{Men2015}, the sound speed in the outer disk is
$c_s \simeq 8\,\mathrm{km\,s^{-1}}$.
This yields a sound-crossing timescale
$t_{\rm sc} \equiv R_{\rm d}/c_s \simeq 20~\mathrm{d}$,
which is more than an order of magnitude shorter than the observed precession period
$P_{\rm prec} = 587$~d.
The disk can therefore efficiently communicate across its radial extent on a timescale much shorter than the precession timescale, satisfying the standard criterion for approximately rigid-body nodal  precession.

We thus conclude that HD~170582 simultaneously satisfies the observed long-cycle amplitude, period ratio, disk size, and dynamical rigidity constraints within a single tilted-disk precession framework, providing a focused and self-consistent test of the model.

\section{Summary}\label{sum}

We investigate whether the long photometric cycles observed in DPVs can arise from nodal precession of a tilted accretion disk driven by the tidal torque of the companion. 
Within a simple analytical framework, we derive testable relations linking the observed long-to-orbital period ratios to the binary mass ratio $q$, the normalized disk size $R_{\rm d}/a$, and the disk tilt $\cos\beta$.
The tilt angle itself can be inferred from observables, including the long-cycle photometric amplitude $\Delta V$, the orbital inclination $i$, and the disk-to-total luminosity ratio $f_{\rm d}$. 
In systems where $\beta$ cannot be independently constrained, its effect on the precession period can be approximately neglected by setting $\cos\beta \simeq 1$, as disk tilts are generally expected to be small in interacting binaries \citep{Ogi2001,Fou2010}.

We further show that a precessing tilted disk naturally produces either sinusoidal or double-hump long-cycle light curves, matching the two morphologies observed in DPVs. 
The two cases are distinguished by a simple geometric criterion: $i+\beta \le 90^\circ$ for sinusoidal modulation and $i+\beta>90^\circ$ for double-hump modulation.

Applying these relations to a sample of DPVs with measured masses and photometric periods, we find that the majority of the inferred disk radii lie below the expected tidal truncation limit \citep{Pac1977,War1995}, indicating that the model-predicted disk sizes are physically reasonable within the context of interacting binaries.
Only one system nominally exceeds the tidal radius, but remains consistent with it within the quoted uncertainty.

We then compare the modeled disk sizes with independently derived disk sizes from the literature.
We find a positive but statistically weak association between them, accompanied by substantial scatter. 
Although the agreement is not universal, a non-negligible subset of systems is consistent with the tilted-disk precession model within observational uncertainties, without any fine-tuning of free parameters. 
The observed dispersion likely reflects a combination of factors, including heterogeneous disk-radius determinations, intrinsic deviations from the rigid-disk approximation, temporal variability in disk structure, and the possibility that long photometric cycles in some systems are driven by mechanisms other than tilted-disk nodal precession.

For the well-studied system HD~170582, where stellar and disk luminosities, orbital inclination, and long-cycle amplitude are all available, we perform a detailed consistency check. 
We show that the inferred disk tilt angle, disk size, precession period, and sound-crossing timescale are mutually consistent within the tilted-disk precession framework, providing a focused observational test of the model.

Overall, our results demonstrate that tidal nodal precession of a tilted accretion disk can account for the observed long-cycle timescales and amplitudes in a subset of DPVs, supporting its role as a viable and potentially important contributor to the long-period phenomenology of these systems, while also underscoring the need for caution when applying the model to individual objects. 
The analytical framework and testable relations developed here provide a quantitative basis for future observational and theoretical studies of DPVs.

We also comment on the stability of tilted accretion disks during nodal precession.
While our analytical expression is formally identical to that of a rigidly precessing ring, it represents a disk-integrated formulation that reduces to the rigid-ring limit under physically reasonable disk parameters \citep{Rom2000}, with the inferred disk size corresponding to the outer disk radius.
In our sample, the inferred disk sizes generally remain below the expected tidal truncation limit. 
For the well-studied system HD~170582, the sound-crossing timescale across the disk is also significantly shorter than the nodal precession period. 
These conditions favor efficient internal communication within the disk and support the ability of a tilted accretion disk to maintain a coherent structure over multiple precession cycles.
More detailed theoretical and numerical studies also support the long-term stability of nodally precessing disks in interacting binaries.
For example, \citet{Ogi2001} showed that tilted accretion disks can undergo stable nodal precession provided that the viscosity parameter is sufficiently large (typically $\alpha \gtrsim 0.1$), the disk tilt remains small, and the disk radius does not exceed the tidal truncation limit.
Numerical simulations by \citet{Fou2010} likewise showed that tilted disks in X-ray binaries preserve a stable geometry and a nearly constant precession period over the entire duration of the simulations, typically spanning several precession periods.
Although these studies included additional effects such as radiation-driven warping, they indicate that tilted disks need not be intrinsically unstable on precession timescales.
Taken together, these considerations suggest that tilted accretion disks in DPVs with sizes smaller than the tidal radius can plausibly sustain coherent nodal precession over multiple cycles, lending physical support to the applicability of the tilted-disk precession model adopted in this work.

\begin{acknowledgments}
We thank the anonymous referee for constructive comments and suggestions that helped improve this work.
This work was supported by the Science Foundation of Yunnan Province (Nos. 202401AS070046 and 202503AP140013), the International Partnership Program of the Chinese Academy of Sciences (No. 020GJHZ2023030GC), and the Yunnan Revitalization Talent Support Program.
\end{acknowledgments}
\bibliography{refs}{}
\bibliographystyle{aasjournal}

\end{CJK*}
\end{document}